\documentclass[11pt, a4paper]{article}

\usepackage[utf8]{inputenc}
\usepackage[english]{babel}
\usepackage{graphicx}
\usepackage{amsmath,amsthm}
\usepackage{amsfonts,amssymb}
\usepackage{natbib}
\usepackage{url,doi}
\usepackage{a4wide}
\usepackage{hyperref}			

\title{Can Core Flows inferred from Geomagnetic Field Models explain the Earth's Dynamo?}

\author{N. Schaeffer$^1$, E. Lora Silva$^{2*}$, M. A. Pais$^{2,3}$ \\[0.5cm] \small
  $^1$ ISTerre, University of Grenoble 1, CNRS, \emph{F-38041} Grenoble, France \\ \small
  $^2$ CITEUC, Geophysical and Astronomical Observatory, University of Coimbra, Portugal \\ \small 
  $^3$ Physics Department, University of Coimbra, \emph{3004-516} Coimbra, Portugal \\ \small
  $^*$ present address: Department of Chemistry, University of Bath, \\ \small
	  Claverton Down, Bath BA2 7AY, United Kingdom \\
}

\begin{document}

\maketitle

\begin{abstract}

We test the ability of large scale velocity fields inferred from geomagnetic secular variation data to produce the global magnetic field of the Earth.
Our kinematic dynamo calculations use quasi-geostrophic (QG) flows inverted from geomagnetic field models which, as such, incorporate flow structures that are Earth-like and may be important for the geodynamo.
Furthermore, the QG hypothesis allows straightforward prolongation of the flow from the core surface to the bulk.
As expected from previous studies, we check that a simple QG flow is not able to sustain the magnetic field against ohmic decay.
Additional complexity is then introduced in the flow, inspired by the action of the Lorentz force.
Indeed, on centenial time-scales, the Lorentz force can balance the Coriolis force and strict quasi-geostrophy may not be the best ansatz.
When our columnar flow is modified to account for the action of the Lorentz force, magnetic field is generated for Elsasser numbers larger than 0.25 and magnetic Reynolds numbers larger than 100.
This suggests that our large scale flow captures the relevant features for the generation of the Earth's magnetic field and that the invisible small scale flow may not be directly involved in this process.
Near the threshold, the resulting magnetic field is dominated by an axial dipole, with some reversed flux patches.
Time-dependence is also considered, derived from principal component analysis applied to the inverted flows.
We find that time periods from 120 to 50 years do not affect the mean growth rate of the kinematic dynamos.
Finally we notice the footprint of the inner-core in the magnetic field generated deep in the bulk of the shell, although we did not include one in our computations.
\end{abstract}


\section{Introduction}

The main Earth's magnetic field and its temporal variations are generated by the motions of liquid metal in the core.
Provided some assumptions are made, it is possible to infer the large scales of the flow at the top of the Earth's core, from observations of the geomagnetic field and its variations with time, which are commonly referred to as Secular Variation (SV).
Because the crustal magnetic field dominates at small spatial scales, the core field is known only for the largest scales (up to spherical harmonic degree 13).
Similarly, because of time varying currents in the magnetosphere and in the ionosphere, the SV produced by core processes can be isolated only up to harmonic degree 12 to 14 \citep[e.g.][]{olsen14}.
Unfortunately, this inherently also limits to relatively large scales the flow we can reconstruct at the top of the core.

Direct numerical simulations of the geodynamo, pioneered by \cite{glatzmaier95} can be tuned to produce magnetic fields that resemble closely the one of the Earth \citep[e.g.][]{christensen10, christensen11, aubert13CE}.
However, the mechanism by which the magnetic field is actually generated in the Earth's core is not clear.
Indeed, the parameter range in which those simulations operate is arguably very far from the one expected in the Earth, and when extrapolating them to realistic Earth parameters, we obtain a picture where the convective flow is important down to small scales of 0.1 to 10 km wide.
Is this hidden small scale flow (both from numerics and from inversions) important for the generation of the magnetic field?
Or is the large scale flow alone responsible for the induction?
To help answer these difficult questions, the modest goal of this paper is to test the capability for dynamo action of the large scale flows inferred from geomagnetic field models.

The importance of the Coriolis force in the core arguably leads to flows that are mainly invariant along the rotation axis, which are referred to as quasi-geostrophic (QG) or columnar flows.
These columnar flows are thought to be relevant at large spatial scales and short time-scales \citep{jault08,gillet11}.
At smaller spatial scales or more importantly at longer time scales, the columns are expected to wither: at such scales the Lorentz and buoyancy forces should arguably be both taken into account.
The effect of the Lorentz force is of particular interest as \cite{sreenivasan11} have shown that the induced magnetic pumping significantly enhances the dynamo action of the flow.

In this study, we use a columnar flow assumption to infer the flow at the core surface.
The reconstruction of the flow inside the whole core is thus straightforward.
We then use this reconstructed flow in a kinematic dynamo code to explore its dynamo capability.
Despite being a much simpler approach than solving the full dynamical system, solving only the induction equation has already proven to be useful \citep[e.g.][]{gubbins08, tobias13, cabanes14b}. Although care must be taken when applying anti-dynamo theorems to bounded flows \citep[see e.g.][]{bachtiar06,li10}, the quasi-two dimensionality of our columnar flows does not \textit{a priori} help dynamo action \citep{jones08lecture}.
Despite this fact, columnar flows have already proven to be capable of dynamo action in the presence of large-scale zonal shears \citep{schaeffer06, guervilly10phd}.
In addition to the large-scale shears, both studies stress the importance of time-variability of the flow for the dynamo action.
As for inverted core flows, they never exhibit dominant large scale shears. To try to boost the dynamo efficiency of columnar flows, we introduce a time-dependence matching inferred core flows and perturbate the flow with the magnetic pumping described by \cite{sreenivasan11}.
As we will show, only the latter actually leads to dynamo action.

Even though our flow  explains an important part of the observed SV, we do not expect the growing dynamo field to match any particular feature of the secular variation as, e.g., the dipole moment decay rate.
The reasons are inherent to the kinematic dynamo problem and are detailed in section \ref{sec:discussion}.

The point we wish to highlight in the present study is that our flow model, obtained through a geomagnetic field model inversion, and including a physically relevant magnetic pumping that does not change the surface flow, is efficient to maintain the magnetic field of the Earth, without the need for contributions from smaller scales.

%

The paper is organized as follow: in the next section the method for obtaining the core flows is described and discussed.
Then we describe the procedure we follow to compute kinematic dynamos from the core flows, and introduce the relevant parameters.
The results are first described in section \ref{sec:results}, and then discussed in section \ref{sec:discussion}.
We end the paper with concluding remarks.






\section{Inverted core flows}	\label{sec:coreflow}

\subsection{Columnar flows}

Columnar flow can be seen as the outcome of competing mechanisms that propagate information inside the core: Alfv\'en waves due to the magnetic tension in the medium and inertial waves due to the restoring effect of Coriolis force.
On short time-scales and large length-scales, these latter waves are quicker in propagating perturbations axially (along planetary vorticity lines) inside the liquid core, as expressed by a small Lehnert number, i.e. the ratio between Alfv\'en wave and inertial wave speeds \citep{jault08, gillet11, nataf15}.

\cite{schaeffer11} pointed out that equatorially anti-symmetric (AS) features can be present inside a spherical rotating container as a result of some anti-symmetric excitation.
When looking for this symmetry in core flows inverted from geomagnetic field models, they further noticed that the AS component tends to prevail for small scale flows \cite[see also][]{gillet11}, which can be understood in terms of a lengthscale-dependent Lehnert number, $\lambda_\ell=B/\ell \Omega \sqrt{\rho \mu_0}$, larger for small-scale structures.

In a recent work, \cite{pais15} applied Principal Component Analysis tools to the  streamfunction $\xi$ of the QG flow inverted from geomagnetic field models.
The `data', which consisted in the values of streamfunction $\xi$ on a regular spatial grid at the core surface, were decomposed into a mean flow plus a linear combination of a small number of spatial patterns multiplied by time-varying coefficients.
This approach is particularly useful in the present context, by providing the means to describe the time-varying flow derived from geomagnetic field model {\it gufm1} over a 150 year time period with a small number of parameters (see section \ref{sec:time}).

Since allowing for an AS component in inverted core flows improved variation of Length-of-Day ($\Delta LOD$) estimations, \cite{schaeffer11} argued in favor of considering both flow symmetries in the inversion.


\subsection{Obtaining the core flow}

\begin{figure*}
\begin{center}
      \begin{minipage}[c]{0.64\textwidth}
      \includegraphics[scale=0.65]{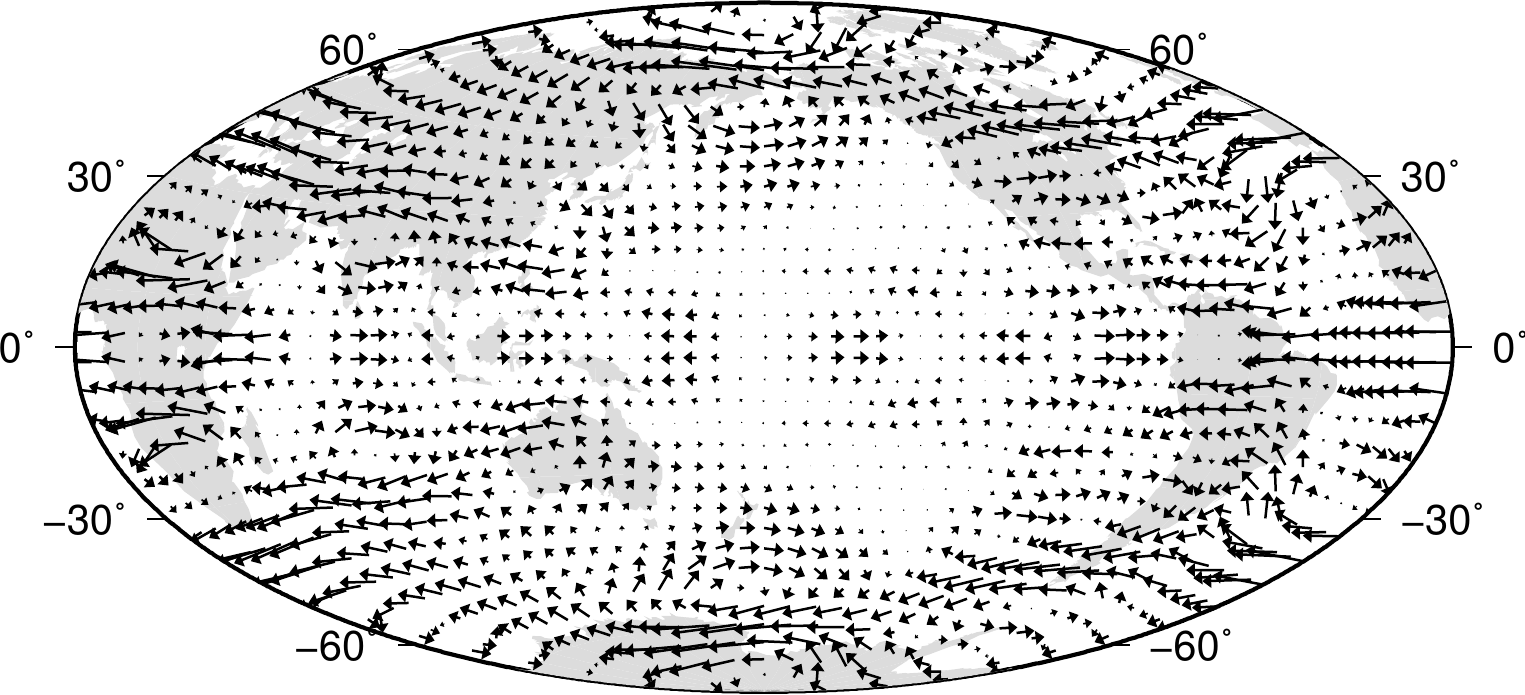}
      \end{minipage} 
      \begin{minipage}[c]{0.32\textwidth}
            \includegraphics[scale=0.55]{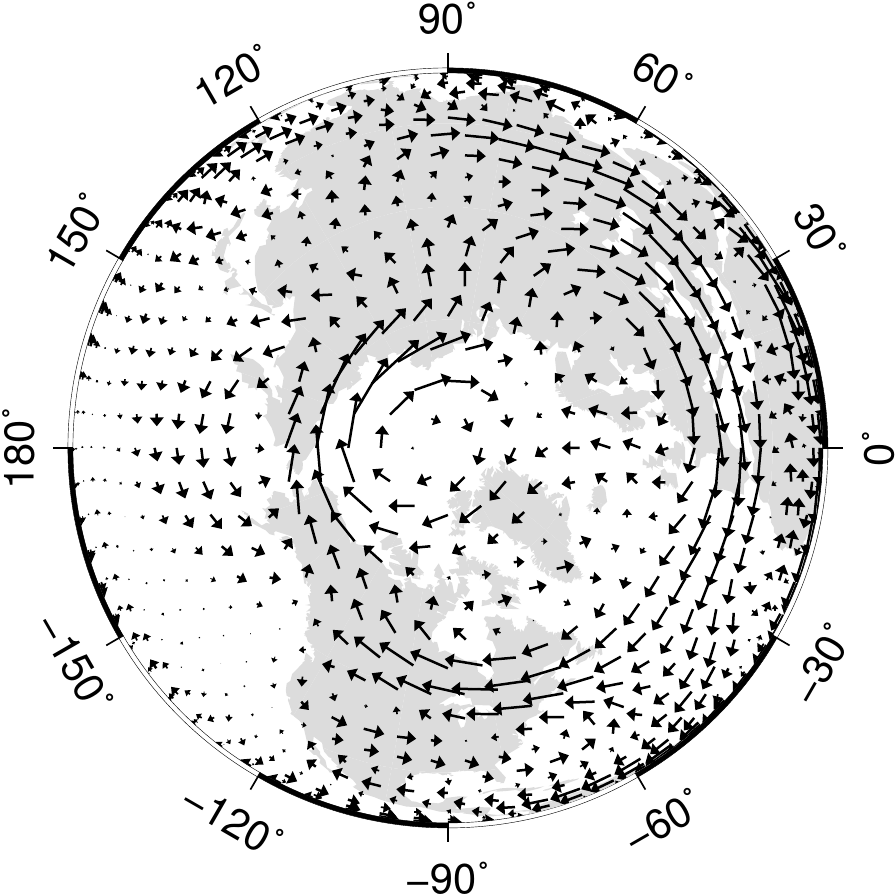}
      \end{minipage} 
      \vspace{-0.3cm}
      \\
      \begin{minipage}[c]{0.32\textwidth}
            \includegraphics[scale=0.55]{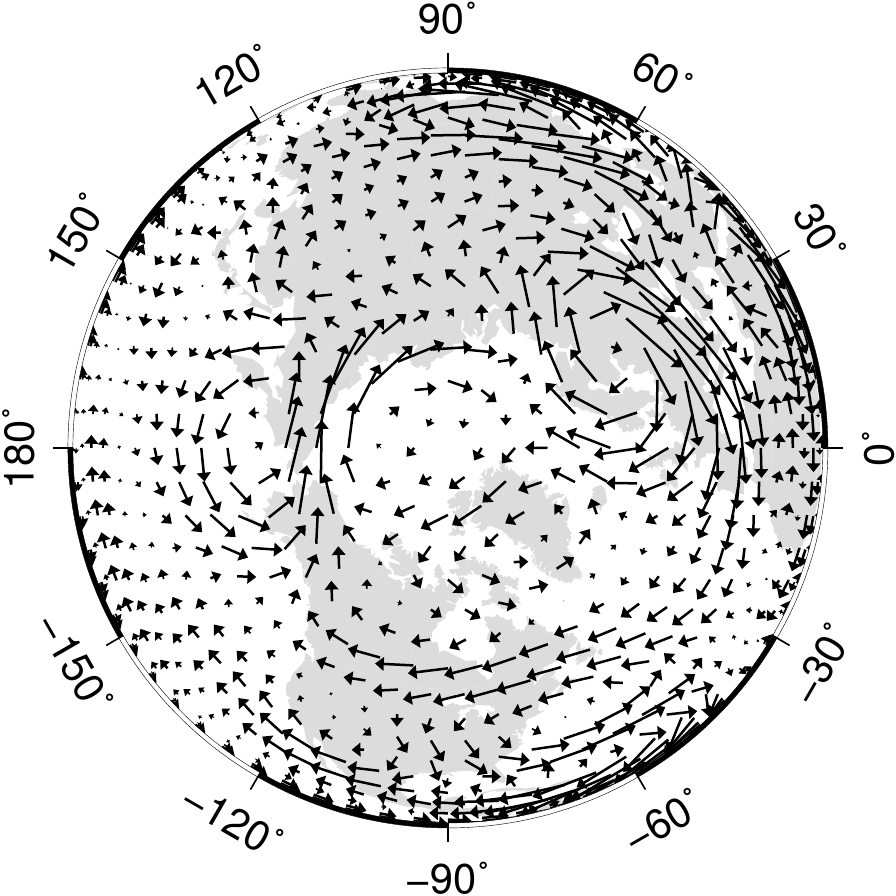}
      \end{minipage} 
      \begin{minipage}[c]{0.32\textwidth}
            \includegraphics[scale=0.55]{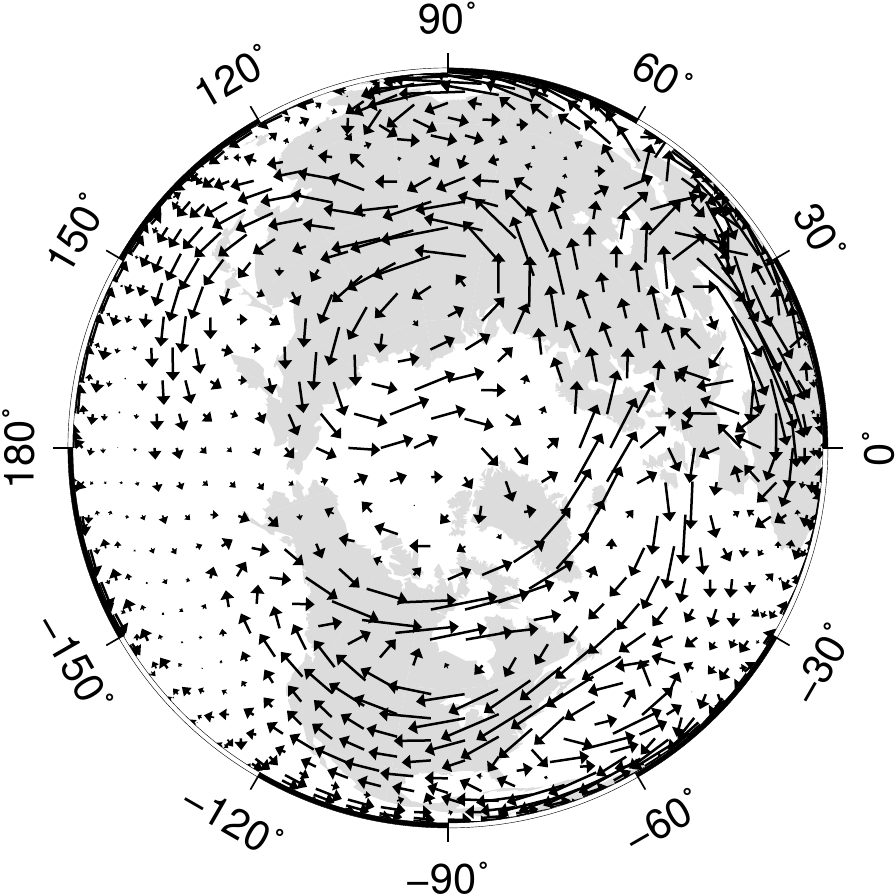}
      \end{minipage} 
            \begin{minipage}[c]{0.32\textwidth}
            \includegraphics[scale=0.55]{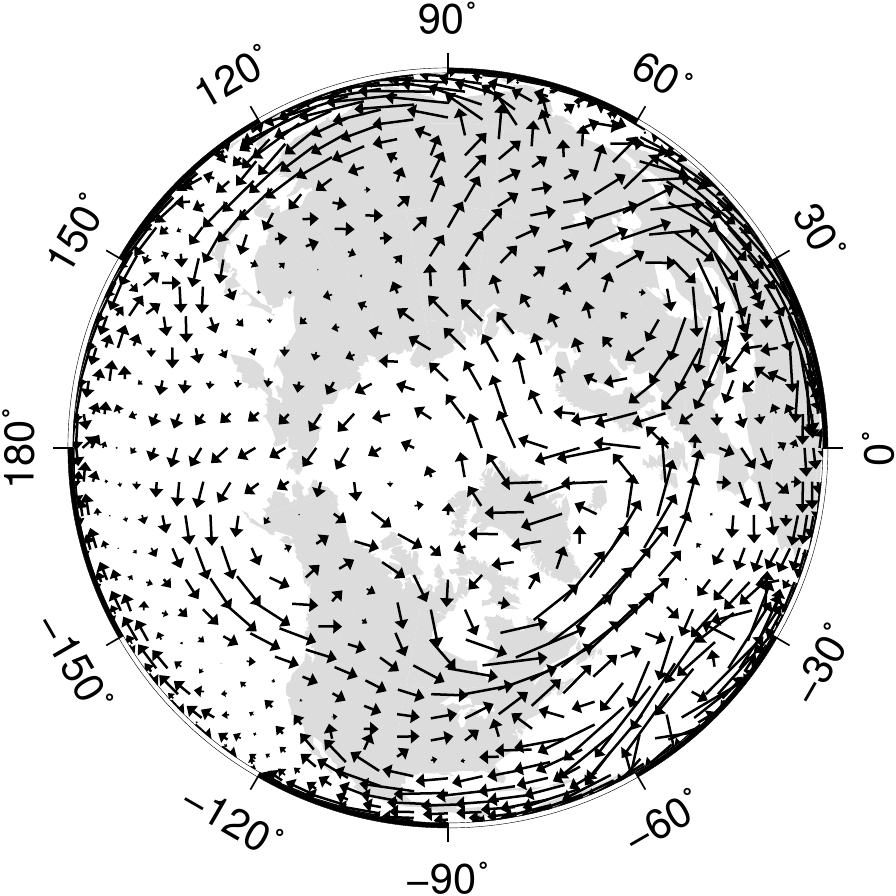}
      \end{minipage} 
 \end{center}
 \caption{On the top row, the mean flow using a Hammer projection centered at the 180$^{\circ}$ meridian (left) and an ortographic projection as seen from the North pole (right). On the bottom row, the three spatial structures or Empirical Orthogonal Functions characterizing the first three modes from the PC analysis. The EOF's are normalised and the scale is arbitrary.} 
\label{fig1}
 \end{figure*}

The geomagnetic field model {\it gufm1} has been inverted for a large-scale columnar flow and an AS component, providing for a flow solution for each epoch in the 1840-1990 period (one independent flow snapshot each year).
The inversion accounts for a separation of QG and AS flows as in \cite{schaeffer11}. 

The columnar flow verifies the following kinematic condition at the core-mantle boundary (CMB):
\protect \begin{equation}
\mathbf{\nabla}_H \cdot \mathbf{u}^+ = \frac{2 \tan \theta}{r_c} u^+_{\theta} \label{QG-flow}
\end{equation}
with $r_c$ the Earth's core radius, whereas the AS component satisfies
\protect \begin{equation}
\mathbf{\nabla}_H \cdot \mathbf{u}^- = 0 \label{tor-flow} \qquad.
\end{equation}
These are the surface constraints of a flow that in the bulk of the core has a QG component, equatorially symmetric (upper index `+'), and an AS component (upper index `-'). 

In \cite{schaeffer11}, the AS component $\mathbf{u}^-$ had no particular kinematical constraint imposed. By using (\ref{tor-flow}) instead, the AS surface flow component can be completely retrieved from a toroidal scalar $\mathcal{T}$, in the same way as the QG surface flow component can be retrieved from a streamfunction scalar $\xi$ \cite[see e.g.][]{pais08, pais15}. 
\protect \begin{eqnarray}
\mathbf{u}^+ &= & \dfrac{1}{\cos \theta} \nabla_H \wedge \xi (\theta, \phi) \, \hat{r} + \dfrac{\sin \theta}{r_c \cos^2 \theta} \, \xi (\theta, \phi) \, \hat{\phi}  \label{eq-xi} \\
\mathbf{u}^- &=  &  \nabla_H \wedge \mathcal{T} (\theta, \phi) \, \hat{r}  \label{eq-tor}  \,
\end{eqnarray}  
where $\xi$ and $\mathcal{T}$ are given in rad/yr.
In this study, where we test the ability of QG flows to increase the energy of magnetic field modes that have a similar morphology to the Earth's field, we extracted $\mathbf{u}^+$ from the inverted flows and did not further consider the AS component $\mathbf{u}^-$.

Besides the two conditions (\ref{QG-flow}) and (\ref{tor-flow}) imposed on the surface flow through quadratic forms on the flow coefficients multiplied by relatively high regularization parameters, two further regularizations are used: a penalization of the azimuthal gradients, $\int_{CMB} [ \left( 1/\sin \theta \right) \left( \partial \mathbf{u} / \partial \phi \right)^2 ] dS$ as in \cite{schaeffer11}, supporting the observation of structures developed preferably along parallels in natural rotating flow systems; a penalization of radial vorticity and horizontal divergence, $\int_{CMB} [ \left( \hat{r} \cdot \nabla_H \wedge \mathbf{u} \right)^2 + \left( \nabla_H \cdot \mathbf{u} \right)^2 ] dS$ , corresponding to the $\ell^3$ norm used in \cite{gillet09} and favoring a large scale flow. 

Following \cite{pais15}, the streamfunction $\xi$ obtained for the QG flow is analyzed into a linear combination of a small number of spatial patterns multiplied by time-varying coefficients, using Principal Component Analysis (PCA).
Five modes are retained, which explain more than $95\%$ of the time variability of the flow.
The reduced streamfunction model reconstructed from these main modes is given by
\begin{equation}
\xi(r_c, \theta, \phi, t )=\xi^0 (r_c, \theta, \phi) + \sum\limits_{k=1}^5 PC_k(t) \, \xi^k_{EOF}(r_c,\theta,\phi)  \label{eq-pc}
\end{equation}
where $\xi^0$ is the mean flow, obtained by averaging the flow coefficients over the time-period 1840 - 1990, $\xi^k_{EOF} $ is the Empirical Orthogonal Function (EOF) of order $k$ depending on spatial coordinates, and $PC_k(t)$  is a function of time, the Principal Component (PC) of order $k$.
The product $PC_k (t) \, \xi^k_{EOF}(r_c,\theta,\phi)$ is the $k$th PCA mode, and explains a percentage of the time variability found in data.
The different modes are uncorrelated in time during the 1840-1990 interval and in space over the whole CMB.
Figure \ref{fig1} shows the flow at the CMB captured by $\xi^0$ and the first three EOFs.
Figure \ref{fig2} shows plots of the five first PCs that enter expression \ref{eq-pc}.
Note the resemblance between this flow and those in \cite{schaeffer11} and \cite{pais15}.
In all cases, the mean flow shows strong azimuthal currents at high latitudes centered at -145$^\circ$ longitude and at low latitudes centered at 0$^\circ$.
The three large vortices at medium/high latitudes described in \cite{pais15} are also present.

The polar anticyclone lying inside the tangent cylinder cannot be conveniently retrieved with conventional QG flow regularization \citep[e.g.][]{pais08}.
It is nonetheless an important feature of core flows which has already been discussed in both observations \citep[e.g.][]{olson99}, and geodynamo simulations \citep[e.g.][]{aubert05}.

\section{A kinematic dynamo problem}

\subsection{Induction equation}

The evolution of the magnetic field within an electrically conducting fluid is given by the induction equation
\begin{equation}
\dfrac{\partial \mathbf{B}}{\partial t} = \mathbf{\nabla} \times \left( \mathbf{v} \times \mathbf{B} \right) + \eta \mathbf{\nabla}^2 \mathbf{B}.	\label{eq-ind}
\end{equation}
where $\eta = (\mu_0 \sigma)^{-1}$ is the magnetic diffusivity, $\mu_0$ is the magnetic permeability of empty space, and $\sigma$ is the electrical conductivity of the fluid that we assume to be homogeneous.
Here, $\mathbf{v}$ denotes the entire three-dimensional flow in the bulk, while $\mathbf{u}$ is the flow at the core surface.

The key parameter for dynamo action is the magnetic Reynolds number
$$
Rm=V_0 r_c/\eta=V_0 r_c \mu_0 \sigma
$$
which compares the magnitude of the induction term to the ohmic dissipation.
We use as characteristic speed $V_0$ the maximum value of the velocity of the flow field $\mathbf{v}$, and as length scale the radius $r_c$ of the core.
The core flows in this study have $V_0 \sim 15$~km/yr, leading to $Rm \sim 800$ to $3000$ for the Earth's core, depending on the estimated electrical conductivity \citep{pozzo12}.

Dynamo action, which is the spontaneous growth of a magnetic field from the motion of a conducting fluid, happens when the induction overcomes ohmic dissipation, which occurs for $Rm > Rm_c$.
Numerical computations are generally needed to determine the critical magnetic Reynolds number $Rm_c$, because it depends on the precise details of the flow.
Efficient flows in spheres have $Rm_c \sim 10$ to $100$ \citep[e.g.][]{dudley89, jones08lecture}.
Note however, that not all flow fields can ultimately trigger dynamo action, in which case $Rm_c$ is not defined.
Finding $Rm_c$ and the fastest growing magnetic field $\mathbf{B}$ when the flow $\mathbf{v}$ is prescribed is referred to as the kinematic dynamo problem.

In the context of geomagnetism, \cite{gubbins08} and \cite{sarson03} studied the kinematic dynamo problem for minimalistic flows that captured some expected features of core flows.
Here, the prescribed velocity field $\mathbf{v}$ and its time evolution are expected to be more Earth-like, since they are derived from the core flow $\mathbf{u^+}$ inverted from geomagnetic field observations as described in section \ref{sec:coreflow}.
This will be detailed below.

\subsection{Numerical procedure}

In order to determine $Rm_c$ for a given flow $\mathbf{v}$, we compute the time-evolution of $\mathbf{B}$ as given by the induction equation (\ref{eq-ind}), for several values of the magnetic diffusivity $\eta$.

We time-step the induction equation (\ref{eq-ind}) numerically, using the XSHELLS code \citep{gillet11, monteux12, cabanes14b}, which is available as free software, and has been adapted for this study \citep[see][]{xs_estelina}.
It uses the spherical harmonic transform library SHTns \citep{schaeffer13} and finite differences in radius.
The integration scheme is second order in time with the diffusion terms treated by a Crank-Nicolson scheme, whereas an Adams-Bashforth one is used for the induction term.

Both the prescribed flow $\mathbf{v}$ and the magnetic field $\mathbf{B}$ occupy the full sphere: we have no solid inner-core in our computations to avoid the issues arising when trying to reconstruct a columnar flow compatible with a solid inner-core.
Note that the XSHELLS code has been benchmarked to full-sphere solutions \citep{marti14} and used in a previous study involving full-spheres \citep{monteux12}.
The mantle is assumed electrically insulating, so that the magnetic field in the liquid sphere matches a potential field outside the conducting domain.

The magnetic energy is monitored, its growth rate indicating if the flow leads to dynamo action.
However, when starting from a random magnetic seed, transient decay and growth may be observed before the average growth-rate converges to a constant value.
The time needed to obtain a well converged mean growth-rate is typically a fraction of the magnetic diffusion time $T_\eta = r_c^2/\eta$.
All our simulations have been running at least for one fifth of $T_\eta$ but often one half of $T_\eta$ and, in a few cases, up to $2 T_\eta$.

A typical run is set up with $N_r=300$ radial levels, and spherical harmonic expansions truncated at degree $\ell_{max}=149$ and order $m_{max}=85$.
For the most demanding cases (largest $Rm$) and to check numerical convergence, we pushed resolution up to $(N_r,\ell_{max},m_{max})=(320,250,150)$.
The computations are always fully dealiased using the appropriate numbers of grid points in latitudinal ($N_\theta > 3\ell_{max}/2$) and longitudinal ($N_\phi > 3m_{max}$) directions.

\subsection{Prescribed flow models}

\subsubsection{Time dependence through PC}		\label{sec:time}

\begin{figure}
\begin{center}
      \includegraphics[width=0.7\columnwidth]{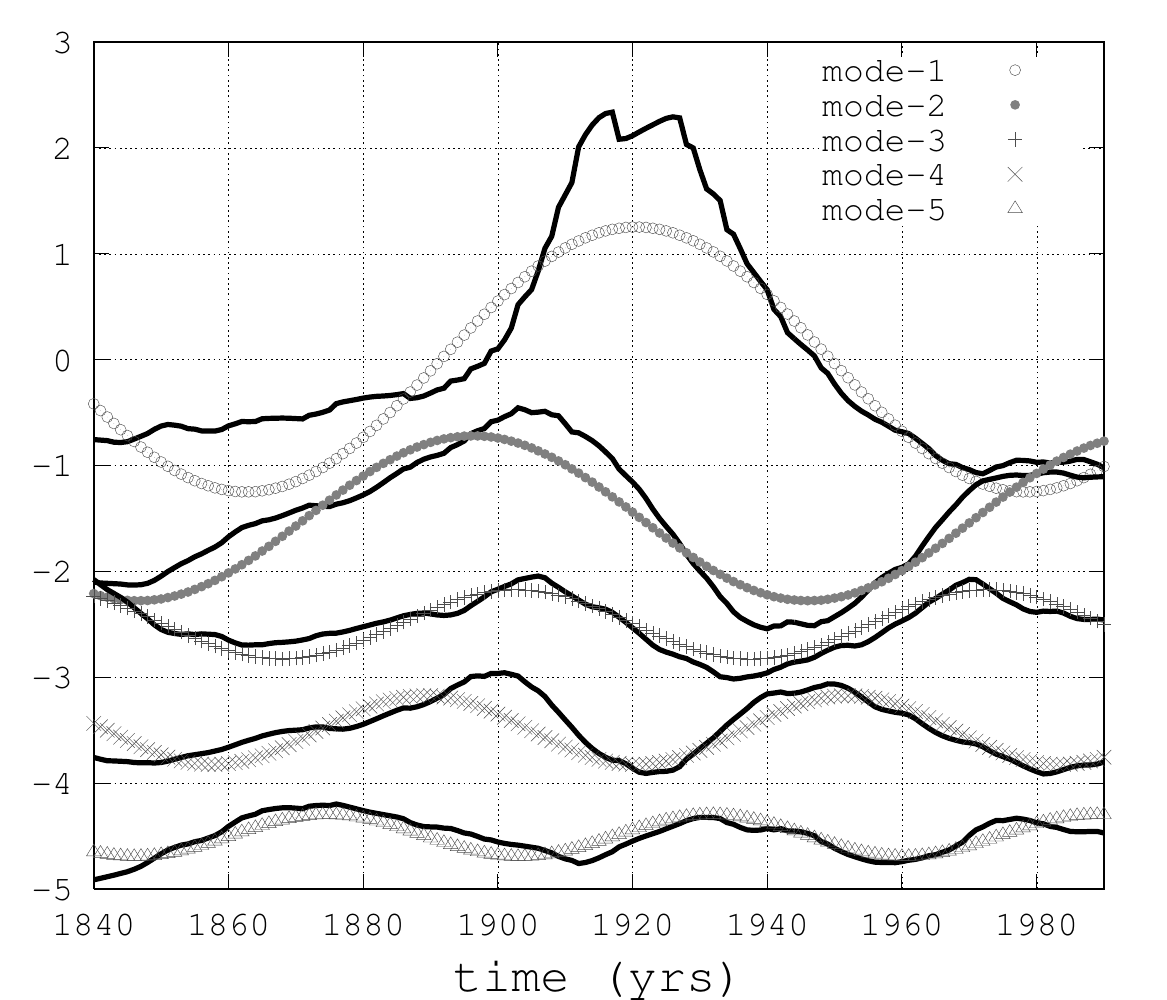}
 \end{center}
 \caption{The main five PCs (used to reconstruct the flow using eq. \ref{eq-pc}) and their fitted sine functions, each defined by an amplitude, a time period and a phase shift.
}
\label{fig2}
 \end{figure}

It has been shown that the time-dependence can be important for dynamo action \citep{willis04, schaeffer06, tilgner08}.
We thus also use flows varying in time in our kinematic dynamo study, as captured by geomagnetic observations.
The prescribed velocity field $\mathbf{u}$ is computed from equation (\ref{eq-xi}), with $\xi(t)$ given by equation (\ref{eq-pc}).
In order to asses if a flow is capable of sustaining a magnetic field against ohmic dissipation, we need to integrate the induction equation (\ref{eq-ind}) for times comparable to the magnetic diffusion time.
Because the magnetic diffusion time in the Earth is much longer than the time-period for which the core flow $\mathbf{u}$ is computed, we cannot reconstruct the flow for a time long enough for a simulation to run.
To overcome this, we fitted each of the first five principal components $PC_k(t)$ (see eq. \ref{eq-pc}) with one sine function $\widetilde{PC}_k(t) = A_k\sin(\omega_k t + \alpha_k)$, as represented in figure \ref{fig2}.
This allows us to compute a flow $\mathbf{u}$ at any time, by extrapolating from decade time scales at which observations take place to much larger diffusive ones at which dynamo action may occur.

The prescribed bulk flow $\mathbf{v}(t)$ is then reconstructed from the symmetric surface flow $\mathbf{u^+}(t)$ using the approach we now describe.

\subsubsection{Purely columnar flow}

The simplest way to reconstruct the flow at any depth in the core, is to assume a columnar flow.
The cylindrical components ($s$,$\phi$,$z$) of the bulk flow $\mathbf{v}$ are related to the spherical components ($\theta$,$\phi$) of the surface flow $\mathbf{u}^+$ by

\begin{eqnarray}
v_s (s, \phi) &=& \dfrac{H(s)}{r_c} u^+_{\theta} (\theta, \phi) \nonumber \\
v_{\phi} (s, \phi)& = & u^+_{\phi} (\theta, \phi) \nonumber \\
v_{z} (s, \phi, z) &=& -\dfrac{s z}{r_c H(s)} \, u^+_{\theta} (\theta, \phi) \label{QG-deep}
\end{eqnarray}
with $s = r_c \sin \theta$ the cylindrical radial coordinate and $H(s)=\sqrt{r_c^2 - s^2}$ the half height of a fluid column.
Because $\mathbf{u}^+$ derives from a pseudo-streamfunction (see eq. \ref{eq-xi}), mass conservation is ensured and we always have $\nabla . \mathbf{v} = 0$ \cite[see also][]{amit13}.
Remember also that in this study, we ignore the AS component $\mathbf{u^-}$.


%
%
%
%
%

\subsubsection{Magnetic pumping}	\label{sec:magpump}

Helicity is another important ingredient for dynamo action as it leads to the so-called alpha-effect \citep[e.g.][]{jones08lecture} whereby poloidal magnetic field is produced from toroidal field.
Ekman pumping in columnar vortices produces helicity \citep{busse75}, but it vanishes at small enough Ekman numbers \citep[e.g.][]{schaeffer06}.

On centenial time-scales the influence of the Lorentz force on the flow should be taken into account.
Indeed, when a magnetic field permeates columnar vortices, \cite{sreenivasan11} have put forward an effect coined magnetic pumping, which produces an axial flow in phase with axial vorticity, that significantly enhances the mean helicity of the flow and consequent dynamo action.
Because fields of dipolar symmetry have a more efficient magnetic pumping, they argued that this effect could explain the preference of dipole-dominated magnetic fields in their simulations.

The magnetic pumping is proportional to the local vorticity and to the square of the magnetic field.
\cite{sreenivasan11} have computed magnetic pumping solutions for a simple toroidal field
\begin{equation}
B_\phi = B_0 \, \frac{s}{r_c} \frac{z (H(s)^2 - z^2)}{H(s)^3}		\label{eq:bphi_dip}
\end{equation}
of dipolar symmetry.

\begin{figure}
\begin{center}
   \includegraphics[width=0.95\columnwidth]{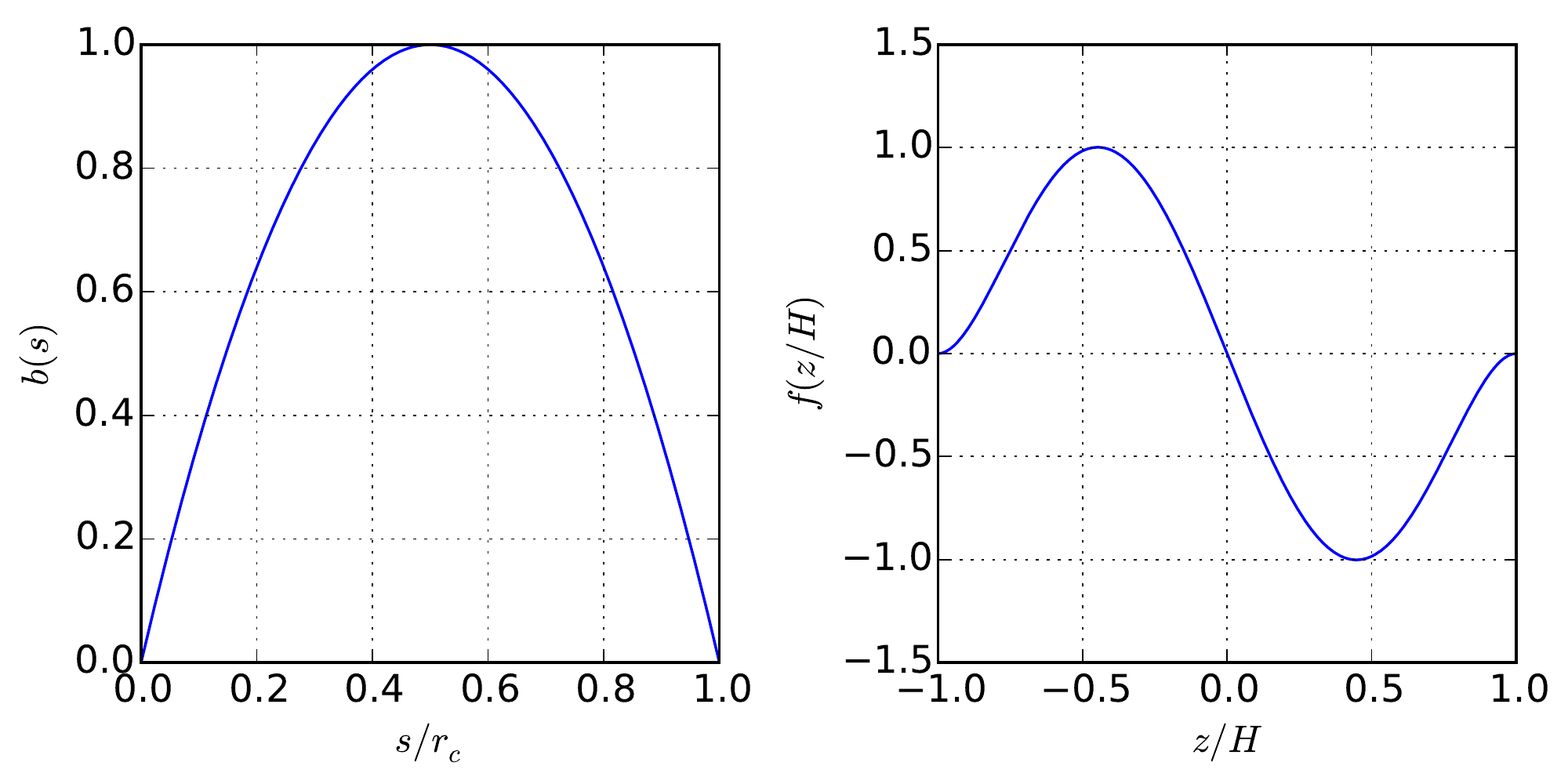}
\end{center}
\caption{Radial (left) and vertical (right) profiles of our parametrized magnetic pumping.
They were chosen to capture the dependence with $s$ and $z$ corresponding to a simple field of dipolar symmetry \cite[see figure 1b of][]{sreenivasan11}.}
\label{fig3}
\end{figure}

Because in our kinematic dynamo approach the flow is prescribed by the streamfunction $\xi$ (eq. \ref{eq-xi} and \ref{eq-pc}), we cannot take into account the retroaction of the dynamic magnetic field.
Instead, we assume the Earth permeated by a simple toroidal field of dipolar symmetry, and follow \cite{sreenivasan11}.
As no explicit expression of the magnetic pumping is available, we choose to model the corresponding axial velocity $v_z^{mp}$ with the following parametrization that closely mimics the magnetic pumping computed by \cite{sreenivasan11} (see their equation 3.9 and their figure 1b).
Using a Fourier decomposition $\xi(s,\phi) = \sum_m \xi_m(s) e^{im\phi}$, we prescribe, for all azimuthal wavenumber $m$:
\begin{equation}
v_z^{mp} = \Lambda \: V_0 \: f(z/H(s)) \, b(s) \: m^2 \xi_m(s) 	\label{eq-magpump1}
\end{equation}
where $f(x) = -\frac{7}{2} x(1-x)^2 (1+x)^2$ captures vertical variations due to the above magnetic field geometry $B_\phi$, and $b(s) = 4s(r_c-s)/r_c^2$ takes into account additional magnetic field variations with $s$.
The profiles $f(x)$ and $b(s)$ are represented in Figure \ref{fig3}, and the assumed toroidal field in Figure \ref{fig6} (left).

The Elsasser number
\begin{equation}
\Lambda = \frac{B_0^2}{\mu_0\rho\Omega\eta}
\end{equation}
controls the strength of the magnetic pumping, where $B_0$ is the maximum of the amplitude of the large scale magnetic field, $\rho$ is the fluid density and $\Omega$ the rotation rate of the Earth. 

Note also that we have approximated the local vorticity by $m^2 \xi_m$, neglecting the contribution of the radial derivative of $\xi (s, \phi)$.
This approximation allows us to conveniently satisfy the mass-conservation by adding a contribution  $v_\phi^{mp}$ to the azimuthal flow:
\begin{equation}
v_\phi^{mp} = \Lambda V_0 \: ism \, \xi_m(s) \, b(s) \, \frac{1}{H(s)} f'(z/H(s))		\label{eq-magpump2}
\end{equation}

Our magnetic pumping flow correction defined by equations (\ref{eq-magpump1}) and (\ref{eq-magpump2}) enhances the helicity of the flow while satisfying the mass conservation as well as the impenetrable boundary condition at the core-mantle boundary.
It is also noteworthy that the magnetic pumping does not alter the surface flow.

Although we could have parameterized the small correction due to buoyancy forces in a similar manner, \cite{sreenivasan11} mention that it has a much smaller effect on helicity.
Therefore, and to keep our study simple, we ignore the small deformation of columns induced by buoyancy effects.

\section{Results}	\label{sec:results}

We have computed kinematic dynamos with a combination of the two parameters $\Lambda$ and $Rm$.
Using only the mean flow $\mathbf{u}^+$ averaged in time and a purely columnar velocity field $\mathbf{v}$ given by equations \ref{QG-deep}, we found no dynamo for any tested value of $Rm$ up to $Rm=2600$.
This is not unexpected as \cite{schaeffer06} also did not find dynamo action with stationary columnar flows.

In addition, adding the time dependence prescribed by our principal component analysis did not help, and no dynamo was found up to $Rm=2600$.
Moreover, we checked that the average growth rate is exactly the same as for the mean flow only.
\cite{willis04} showed that adding low frequency perturbations to a mean flow did not change the growth rate, while higher frequency often enhanced dynamo action.
Since our principal components have periods larger than 60 years, it does not exclude that lower period motions can participate in the dynamo process.

\begin{figure*}
   \begin{center}
   \includegraphics[width=0.32\textwidth]{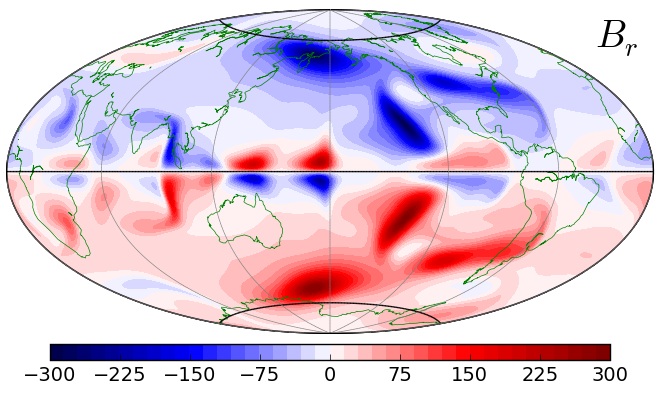}
   \includegraphics[width=0.32\textwidth]{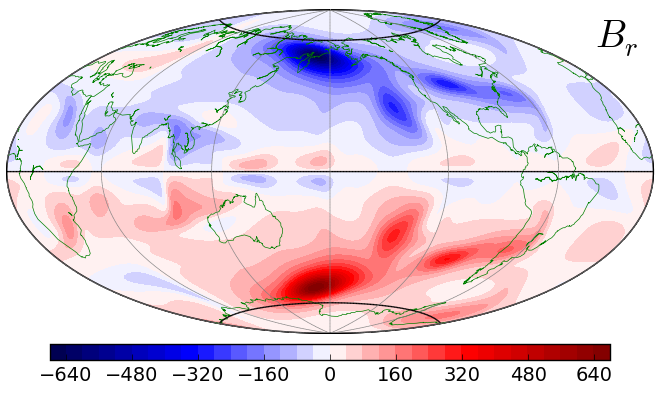}
   \includegraphics[width=0.32\textwidth]{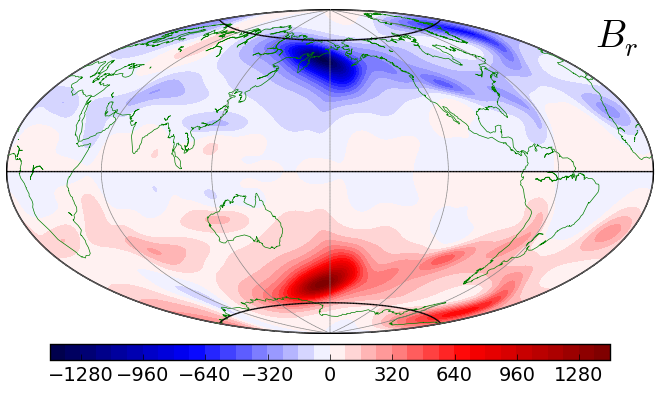} \\
   \includegraphics[width=0.32\textwidth]{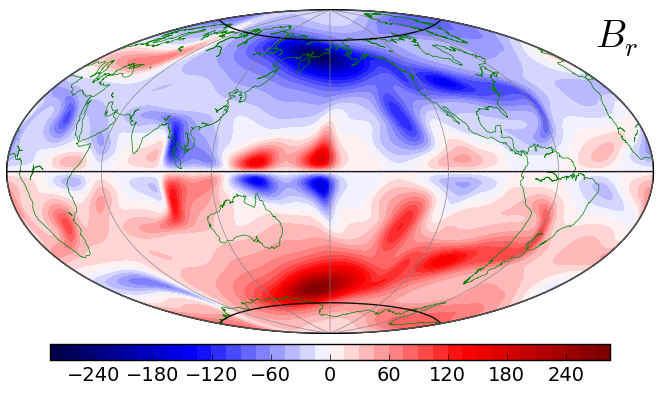}
   \includegraphics[width=0.32\textwidth]{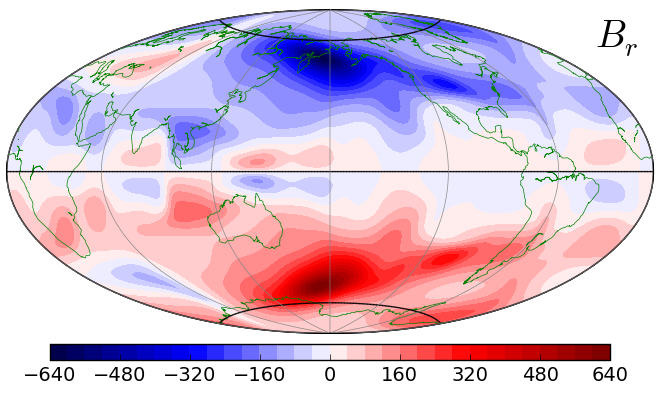}
   \includegraphics[width=0.32\textwidth]{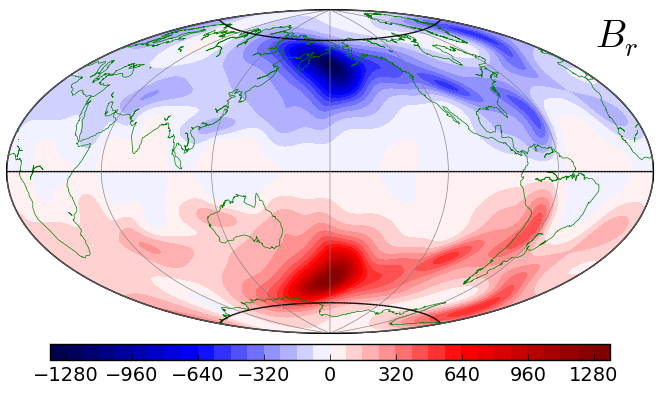}
   \end{center}
   \caption{Growing radial magnetic field shown at the top of the core (Aitoff projection). It has been rescaled to match the assumed Elsasser number $\Lambda = 0.5$ (left), $\Lambda = 0.9$ (middle), $\Lambda=2$ (right).
   For each $\Lambda$, the top row is the largest $Rm$ whereas the bottom row is the smallest $Rm$ leading to a growing field (see table \ref{tab1} for precise values).
   The units are $\mu$T}
   \label{fig:br}
\end{figure*}

\begin{table}
\begin{center}
\begin{tabular}{ccc|cc}
$\Lambda$ & $Rm$ & $Rm^*$ & $\gamma$ & $B_{rms}/B_{dip}$ \\
\hline
0.15 & 1727 & 600 & -7.6 &  \\
0.25 & 1733 & 602 & -0.4 &  \\
0.3 & 1303 & 453 & -1.28 &  \\
0.3 & 1502 & 522 & 1.15 & 18.1 \\
0.3 & 1737 & 604 & 4.44 & 17.2 \\
0.4 & 915 & 318 & -1.5 &  \\
0.4 & 1107 & 385 & 1.5 & 13.6 \\
0.4 & 1308 & 455 & 5.3 & 12.9 \\
0.4 & 1744 & 606 & 14.9 & 14.7 \\
0.5 & 800 & 279 & -0.01 &  \\
0.5 & 700 & 244 & -1.7 &  \\
0.5 & 1000 & 348 & 4 & 11.2 \\
0.5 & 1400 & 488 & 14.5 & 12.1 \\
0.5 & 1751 & 610 & 27 & 14.0 \\
0.7 & 503 & 176 & -2 &  \\
0.7 & 596 & 209 & -0.08 &  \\
0.7 & 706 & 247 & 2.55 & 8.4 \\
0.7 & 1059 & 371 & 12.7 & 10.1 \\
0.9 & 400 & 142 & -1 &  \\
0.9 & 489 & 173 & 0.74 & 6.5 \\
0.9 & 649 & 230 & 4.66 & 7.4 \\
0.9 & 978 & 346 & 14 & 9.2 \\
1.2 & 252 & 91.2 & -1.9 &  \\
1.2 & 360 & 130 & 2.3 & 5.5 \\
1.2 & 495 & 179 & 6.4 & 6.4 \\
1.5 & 201 & 74.7 & -2.9 &  \\
1.5 & 222 & 82.5 & -1.6 &  \\
1.5 & 274 & 102 & 1.4 & 5.1 \\
1.5 & 411 & 153 & 8 & 6.2 \\
2 & 222 & 80.7 & -1.1 & \\
2 & 252 & 92 & 0.9 & 5.2 \\
2 & 454 & 165 & 11 & 6.9
\end{tabular}
\end{center}
\caption{Kinematic dynamo runs using the mean core flow and our magnetic pumping parameterization.
The magnetic Reynolds number is computed using the maximum velocity ($Rm$) or the rms velocity ($Rm^*$).
$\gamma$ is the growth rate in magnetic diffusion time units, computed from the temporal evolution of the magnetic energy $E_B(t)$ as $\gamma = (2E_B)^{-1} \, \partial_t E_B$.
For the growing magnetic field cases ($\gamma>0$), the amplitude ratio between the root-mean-square field in the bulk and the dipole at the surface is given by $B_{rms}/B_{dip}$. 
}
\label{tab1}
\end{table}

\begin{figure}
   \begin{center}
   \includegraphics[width=0.7\columnwidth]{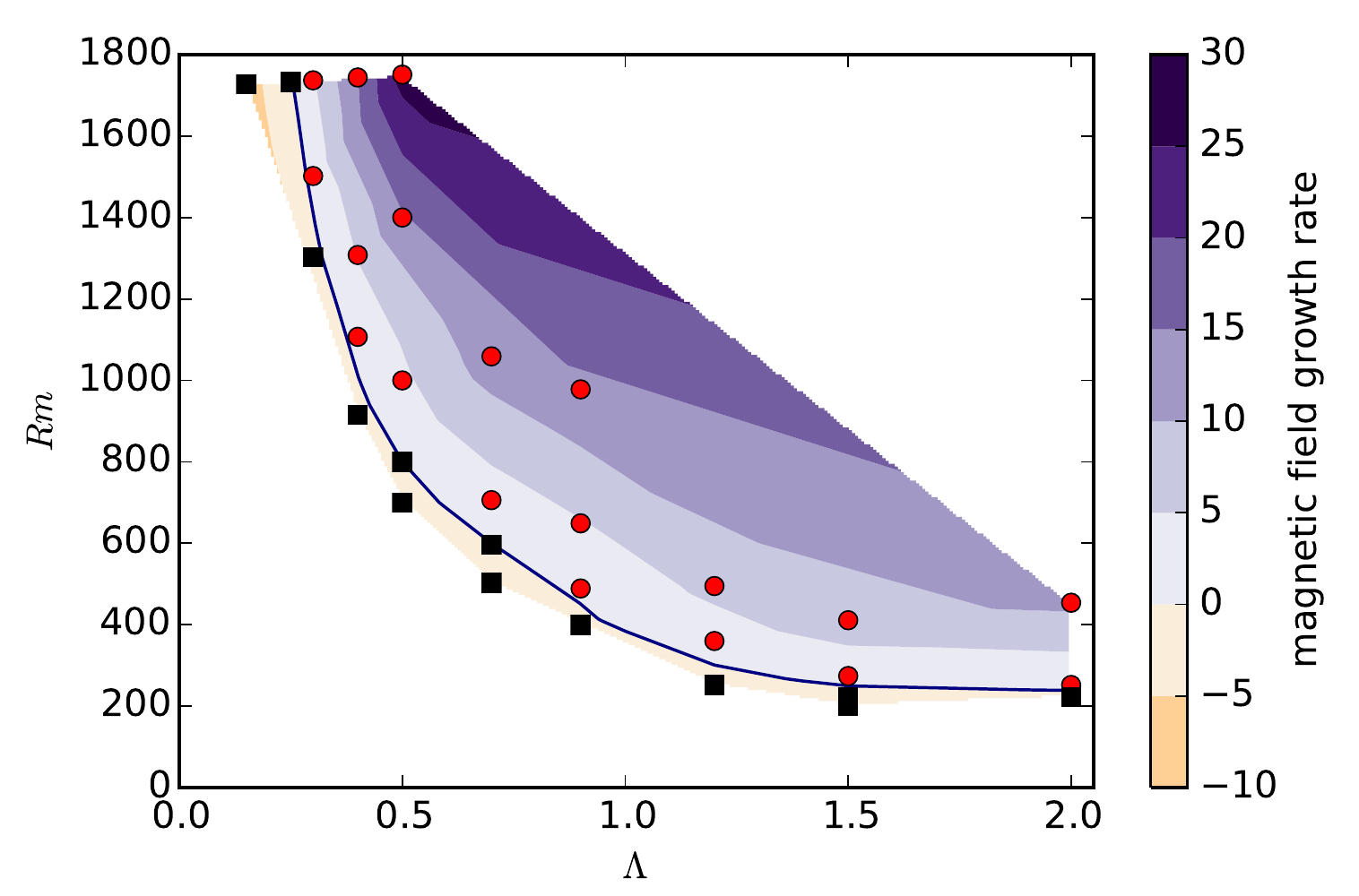}
   \end{center}
   \caption{Growth rate as a function of the magnetic Reynolds number $Rm$ and the Elsasser number $\Lambda$ controlling the magnetic pumping.
   Black squares are failed dynamos (for which the magnetic field decays) while red circles are dynamos (exhibiting growing magnetic field). The contour lines are interpolated using the shown data points.
   The growth rate has been normalized using the magnetic diffusion time.
   The data is given in table \ref{tab1}.
   }
   \label{fig:rm_vs_lambda}
\end{figure}

Nevertheless, as an auspicious result, adding magnetic pumping (eq. \ref{eq-magpump1} and \ref{eq-magpump2}) to our core flows allowed them to produce growing magnetic fields:
the larger the Elsasser number $\Lambda$, the lower the critical magnetic Reynolds number, down to $Rm_c \sim 250$.
Below $\Lambda = 0.5$, we could not find a dynamo for $Rm$ up to 1700.
These computations are summarized in table \ref{tab1} and figure \ref{fig:rm_vs_lambda}, showing a regime diagram for dynamo action with magnetic pumping.
We note that the time-dependence has still no effect on the average growth rate, even in the presence of magnetic pumping.

Since our flow $\mathbf{v}$ is symmetric with respect to the equatorial plane, the magnetic field eigen modes of the kinematic dynamo problem also have a definite symmetry (either symmetric or anti-symmetric).
It turns out that the fastest growing magnetic field is anti-symmetric with a large dipolar component (see figure \ref{fig:br}), which is consistent with the assumption used to model the magnetic pumping (see section \ref{sec:magpump}).

\begin{figure}
   \begin{center}
   \includegraphics[width=0.7\columnwidth]{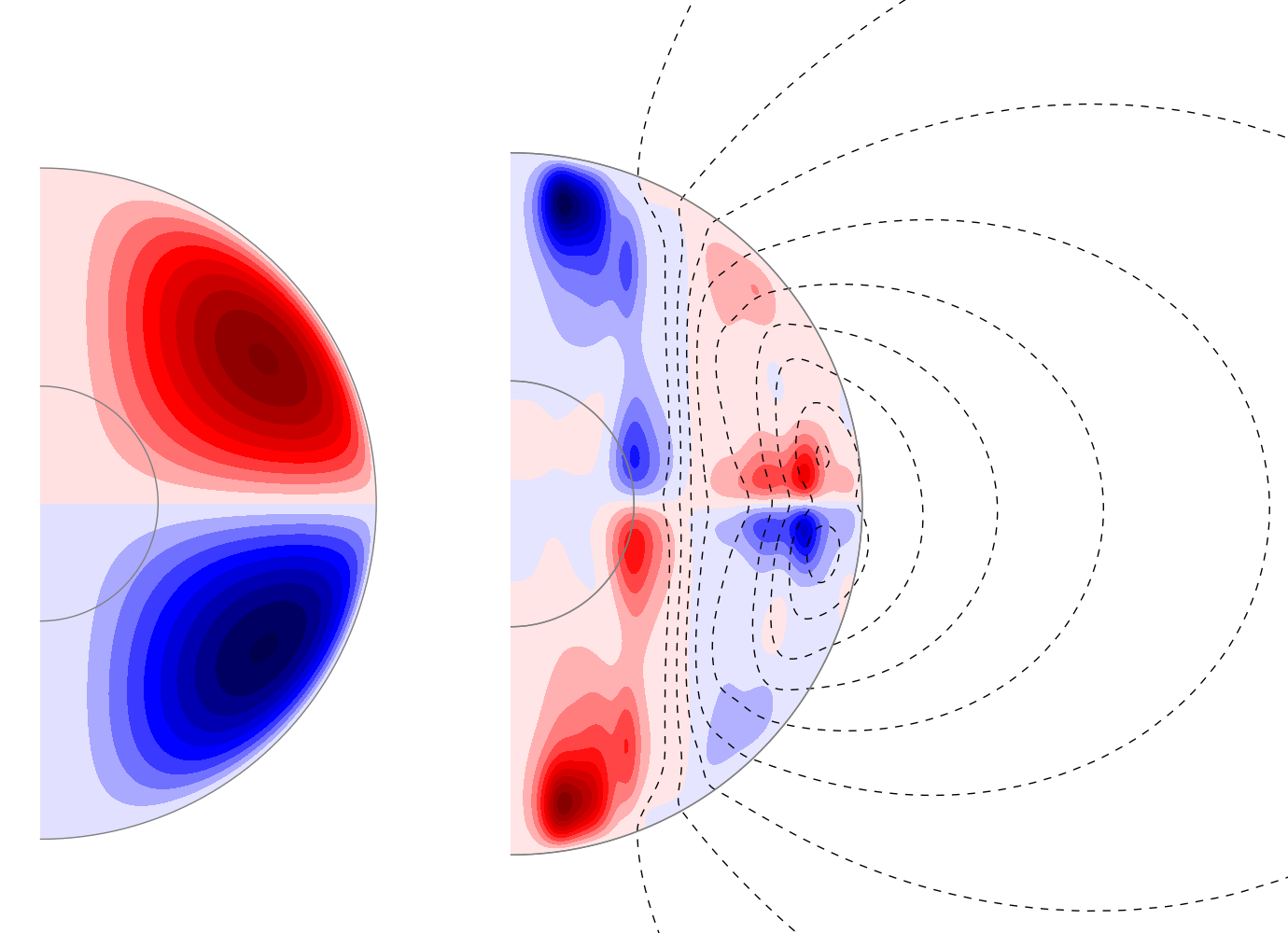}
   \end{center}
   \caption{Left: the simple magnetic field used in eqs. \ref{eq-magpump1} and \ref{eq-magpump2} to model the magnetic pumping.
   Right: axisymmetric part of the growing magnetic field for $Rm=978$ and $\Lambda=0.9$.
   The colormap shows the azimuthal (toroidal) component and the black contours show the meridional (poloidal) field lines.
   The small gray circle marks the size of Earth's inner-core, although we have none in our study.}
   \label{fig6}
\end{figure}

Because the growing magnetic field is likely different from the simple assumed toroidal field responsible for the magnetic pumping in our framework, the resulting kinematic dynamos are not \textit{a priori} dynamically self-consistent.
As an \textit{a posteriori} consistency check, we can compare the growing axisymmetric toroidal magnetic field to the one we have used to compute the magnetic pumping in equations \ref{eq-magpump1} and \ref{eq-magpump2}.
As shown in Fig. \ref{fig6}, they are in fair agreement, although more complexity can be seen in the growing magnetic field.
In particular they are both equatorially antisymmetric.

\section{Discussion}	\label{sec:discussion}

While \cite{guervilly12} used a forcing at the surface to produce a dynamic bulk flow compatible with the observed zonal jets of giant planets, here we decided to entirely prescribe the flow, as the core flow is more complex and time dependent.

\begin{figure}
   \begin{center}
   \includegraphics[width=0.7\columnwidth]{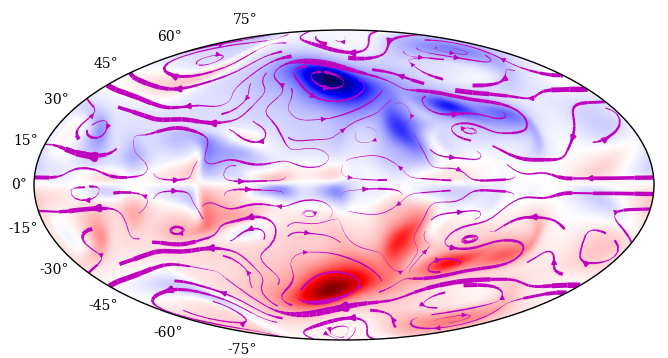}
   \includegraphics[width=0.7\columnwidth]{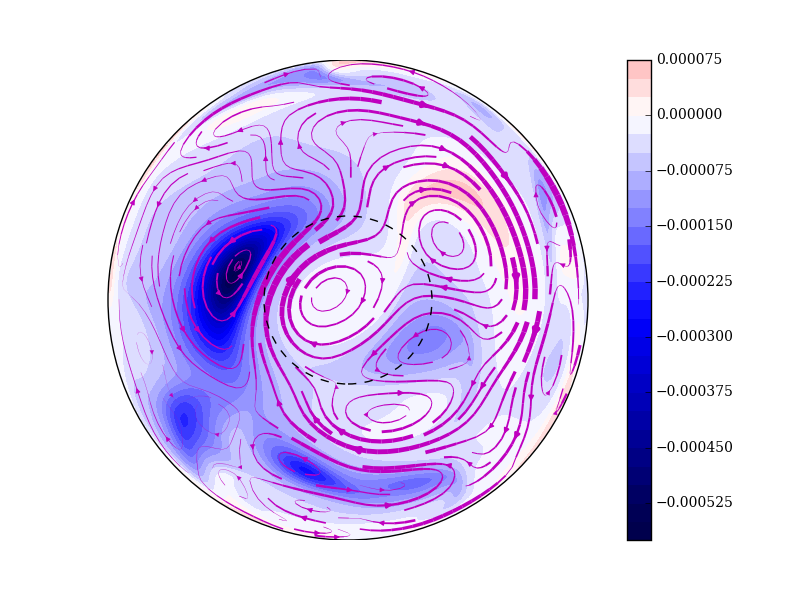}
   \end{center}
   \caption{Growing radial magnetic field (colormap) and streamlines of the surface mean flow for $Rm=978$ and $\Lambda=0.9$.
   Top: Aitoff projection at the core surface centered on the pacific;
   bottom: north-pole view projected onto the equatorial plane (the Greenwich meridian is on the right, as in Fig. \ref{fig1}).
   The thickness of the streamlines is proportional to the velocity. The dashed circle marks the boundary of the Earth's inner-core.}
   \label{fig7}
\end{figure}

We have observed growing magnetic fields in columnar flows with  magnetic pumping, for magnetic Reynolds number $Rm=V_0r_c/\eta$ as low as 252 (for Elsasser number $\Lambda=2$).
If we use the root-mean-square velocity in the bulk instead of the maximum velocity $V_0$, it translates to $Rm^* = 92$.
This rather low value of $Rm^*$ indicates that magnetic pumping in columnar flows is a rather efficient mechanism for dynamo action. 

Our flow is a model of the large scale flow in the Earth's core where, because the operating dynamo is saturated, the Lorentz force backreacts on the flow to prevent further exponential growth of the magnetic field.
Flows from such saturated dynamos have been found to still be efficient kinematic dynamos \citep{cattaneo09,tilgner08b}.
If our flow is mainly responsible for the saturated geodynamo, then according to \cite{cattaneo09} it is also a kinematic dynamo for a growing magnetic field not everywhere proportional to the geomagnetic field and not backreacting on the flow (a passive field).
Our kinematic dynamo calculations thus exhibit the property of velocity fields arising from a saturated dynamo to lead to exponential growth of a passive magnetic field.

Furthermore, in a kinematic dynamo, the obtained field is an eigenmode, growing or decaying at the same rate everywhere, effectively enslaving the time variation of the surface field to the time variation in the bulk.
It may be worth emphasizing that because of vanishing radial velocities at the core surface, the magnetic field there is connected to the bulk field only through magnetic diffusion. As a result, in a regime of exponentially growing field, the dipole moment is also expected to grow.
In contrast, the present geomagnetic field is expectedly in a saturated regime, where Lorentz forces play an important role back-reacting on the flow, although diffusion has probably also a non negligible effect.
In particular, the observed rapid dipole moment decrease has been ascribed to either growth of reverse geomagnetic flux patches in the Southern Hemisphere \citep[e.g.][]{gubbins06}, or advection of normal flux to the equator and of reversed flux to the poles \citep[e.g.][]{olson06}.
Hence, particular features of the observed SV are much dependent on the particular geodynamo regime and cannot be considered here.
In particular, we cannot compare the growth rate obtained for our passive field with the current decay rate of the geomagnetic (saturated) field strength.
In fact, the present decay of the zonal dipole magnetic field energy which is being observed for the last 170 years, could in principle fit into the time variability observed in saturated geodynamos \citep[see e.g.][fig. 6]{christensen11}.

\begin{figure}
   \begin{center}
   \includegraphics[width=0.7\columnwidth]{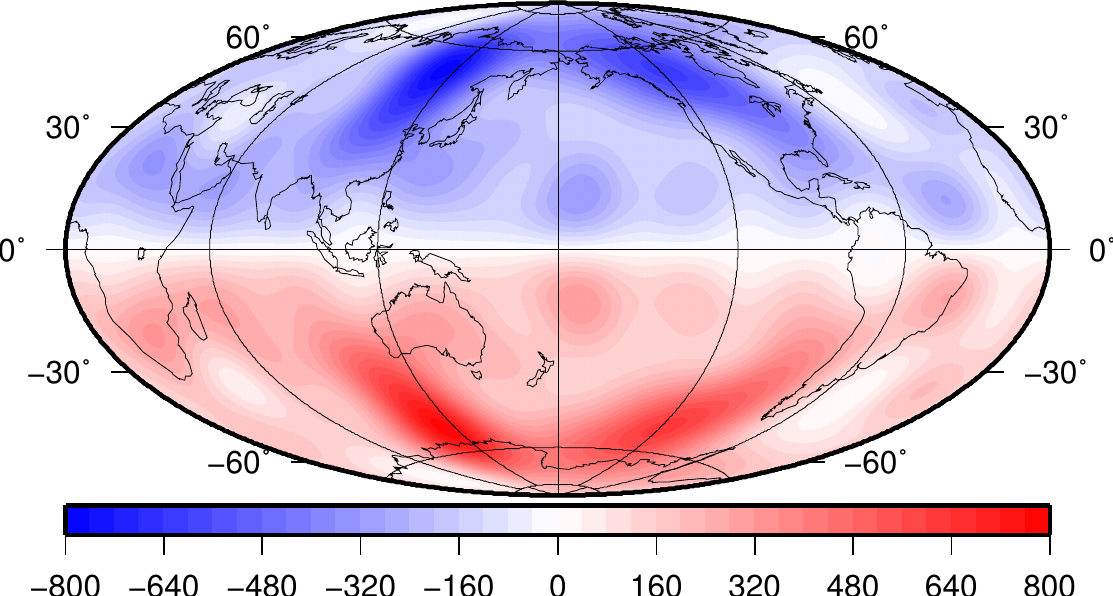}
   \end{center}
   \caption{Anti-symmetric part of the radial magnetic field of \textit{gufm1}, averaged from 1840 to 1990 (Hammer projection).
   Units are in $\mu$T.}
   \label{fig-Br_mean}
\end{figure}

\citet{cattaneo09} have also shown that, in spite of their differences, the growing passive field and the self-consistent saturated field have similar spatial structures.
It may thus be interesting to compare the spatial structure of the growing magnetic field that emerges from our kinematic dynamos to the Earth's internal magnetic field observed nowadays.
To ease this comparison, the mean anti-symmetric magnetic field of \textit{gufm1} is shown in figure \ref{fig-Br_mean}.
The morphology of the growing field does not vary too much within the portion of the parameter space we have explored (see Fig. \ref{fig:br}).
It is always of dipolar symmetry, consistent with the simple toroidal field assumed to compute the magnetic pumping (eq \ref{eq-magpump1}).
As a general trend, larger $Rm$ have more pronounced small-scale features, and large $\Lambda$ lead to fewer but stronger intense flux patches, and also lower amplitude reverse flux patches near the equator.
A noticeable feature is the strong flux patch located right under the Bering Strait, which is associated with a large vortex in the mean flow, as can be seen in figure \ref{fig7}.

We want to emphasize that a perfect match with the Earth's magnetic field is not expected.
Indeed, our mean flow computed over 150 years is unlikely to be an accurate estimate of the mean flow over a magnetic diffusion time (about $10^5$~yrs).
Hence, the strong magnetic flux patch under the Bering Strait, clearly associated with an intense vortex in the mean flow as seen in figure \ref{fig7} and having no corresponding feature in the present Earth's field, may indicate that this very vortex is not part of the long-term mean flow.
However, the presence of symmetric flux patches in the Earth's field not far from this location (with northern hemisphere signatures under North America and Eastern Russia, fig. \ref{fig-Br_mean}), may also suggest that the whirl structure has suffered some displacement around its present position, over very long time periods.
The fact that this vortex is also present in the first EOF (see figure \ref{fig1}) supports our hypothesis.

It is also satisfying that, when rescaling the magnetic field to match the imposed $\Lambda$, the radial field at the top of the core has the same order of magnitude as the Earth's field. Furthermore, the amplitude ratio between the root-mean-square field in the bulk and the dipolar field at the surface is around $\sim 10$ (see table \ref{tab1}), in agreement with numerical geodynamo models \citep{christensen11} and torsional oscillations of $\sim$ 6 yrs periodicity retrieved from geomagnetic field models \citep{gillet10}.


Another striking result is the footprint of the inner-core in the magnetic field generated deep inside the sphere, in spite of no inner core being present in our numerical computations.
This display of the inner core presence is manifest in the change of sign of the magnetic field near the (virtual) tangent cylinder (see figure \ref {fig6}).
This is related to the visible dichotomy in the prescribed flows (see figures \ref{fig1} and \ref{fig7}).

The mechanism of dynamo generation that is considered in this study involves helicity.
Here, an enhancement of helicity is prescribed in the bulk, which propagates no footprint whatsoever up to the core surface.
It is, accordingly, contrasting with the scenario considered by \cite{amit04} where the in-phase occurrence of vorticity and flow shear in the  volume is assumed to manifest on the core surface in the form of correlated horizontal divergence and radial vorticity.

For $\Lambda = 0.25$ and below, no dynamos were found for $Rm \leq 1700$, as shown in figure \ref{fig:rm_vs_lambda}.
It appears that below some magnetic field strength, dynamo action due to the magnetic-pumping is lost, meaning that a strong magnetic field is needed for this mechanism to work.
This is known as a subcritical behaviour \citep{sreenivasan11, morin09}.
Our findings thus suggest that today, a subcritical dynamo may actually produce the magnetic field of our planet.
If this is the case, a sudden change in our planet's field may occur when in the distant future the magnetic field bulk intensity drops below some threshold.

The magnetic-pumping model used in this study is admittedly crude.
A more realistic model, which would dynamically adjust the magnetic-pumping to the actual magnetic field, would allow more realistic kinematic dynamos, with dynamic adjustment of the pumping flow to the growing magnetic field.
It may also be possible to improve the quasi-geostrophic dynamos of \cite{schaeffer06} and \cite{guervilly10phd} towards non-linear quasi-geostrophic geodynamo models.

\section{Conclusion}

We have shown that magnetic pumping in columnar core flows, leads to dynamo action.
The distinctive feature of our study is to use a prescribed velocity field and its time evolution, derived from current geomagnetic field observations.
This expectedly gives them an Earth-like nature, exhibiting features with specific geographical location and vorticity distribution that are most relevant to test their dynamo action.
Note also that our kinematic dynamo approach allows us to reach realistic values of the electrical conductivity.

Furthermore, we show that relaxing the strict axial invariance with a magnetic-pumping is enough for large scale core flows to produce a dipolar field that resembles the one of the Earth, without the contribution of small scales.
As an additional consistency check, the magnetic field in the bulk is about 10 times larger than at the surface (see table \ref{tab1}), in agreement with the study of \cite{gillet10}.
Magnetic pumping is arguably an important mechanism for the geodynamo.

%

In our computations, the magnetic field generation takes place in the bulk of the core where the flow is perturbated, and the surface magnetic field is connected to the bulk field only by diffusion.
In the Earth's core, in a saturated but fluctuating regime, it is still unclear how much magnetic diffusion can contribute to the SV.
Studies bridging numerical dynamo simulations and geomagnetic field inversions may help to shed more light on the subject, by providing estimates of the radial diffusion at the core surface.

For about fifty years, time-dependent geomagnetic field models have been used to infer the geometry and intensity of large scale flows in the core responsible for secular variation.
Are these flow features expected to be present at diffusive timescales, much larger than the time window where we can constrain them with observations?
The answer depends also on their ability for dynamo action.
Our present results, together with previous studies proposing a possible dynamical mechanism for the main mean flow features such as the large eccentric jet \citep[see][]{aubert13CE}, suggest indeed that they could be present for a very long time.

\section*{Acknowledgments}
The authors thank A. Jackson, R. Holme and an anonymous reviewer for their useful comments.
E. Lora Silva and M. A. Pais were supported by FCT (PTDC/CTE-GIX/119967/2010) through the project COMPETE (FCOMP-01-0124-FEDER-019978).
M. A. Pais is grateful to the Grenoble University for funding her stay as an invited professor for two months.
ISTerre is part of Labex OSUG@2020 (ANR10 LABX56).
Most of the computations were performed using the Froggy platform of the CIMENT infrastructure (\texttt{https://ciment.ujf-grenoble.fr}), supported by the Rh\^ one-Alpes region (GRANT CPER07\_13 CIRA), the OSUG@2020 labex (reference ANR10 LABX56) and the Equip@Meso project (reference ANR-10-EQPX-29-01).
The colorful figures were plotted using matplotlib (\texttt{http://matplotlib.org/}).

\bibliography{coimbra}

\appendix

\end{document}